\documentclass[11pt]{article}
\usepackage{times}
\usepackage{mathfont}

\input epsf
\def\efig#1#2{\hbox{\epsfxsize=#1\epsfbox{#2}}}

\newtheorem{theorem}{Theorem}
\newtheorem{lemma}{Lemma}

\DeclareSymbolFont{lasy}{U}{lasy}{m}{n}
\let\Box\undefined
\DeclareMathSymbol\Box{0}{lasy}{"32}
\newcommand{\qed}{\hfill$\Box$}
\newenvironment{proof}{\noindent{\bf Proof:}}{\qed\medskip}

\makeatletter

\long\def\@makecaption#1#2{
   \vskip 10pt 
   \setbox\@tempboxa\hbox{{\small #1. #2}}
   \ifdim \wd\@tempboxa >\hsize   % IF longer than one line:
       {\small #1. #2}\par        %   THEN set as ordinary paragraph.
     \else                        %   ELSE  center.
       \hbox to\hsize{\hfil\box\@tempboxa\hfil}  
   \fi}

\def\@begintheorem#1#2{\it\trivlist
  \item[\hskip\labelsep{\bf #1\ #2.\ }]}
\def\@opargbegintheorem#1#2#3{\it\trivlist
  \item[\hskip\labelsep{\bf #1\ #2\ {\rm(#3)}.}]}

\makeatother

\begin{document}
\bibliographystyle{abuser}

\title{Algorithms for Coloring Quadtrees}

\author{David Eppstein\thanks{Dept.\ of Information and
Computer Science, Univ.\ of California, Irvine, CA 92697-3425, USA,
eppstein@ics.uci.edu, http://www.ics.uci.edu/$\sim$eppstein/.  Work
supported in part by NSF grant CCR-9258355 and by Xerox Corp.}
\and Marshall W. Bern\thanks{Xerox Palo Alto Research Ctr., 3333 Coyote
Hill Rd., Palo Alto, CA 94304, USA, bern@parc.xerox.com.}
\and Brad Hutchings\thanks{Dept.\ of Information and
Computer Science, Univ.\ of California, Irvine, CA 92697-3425, USA.}}

\date{}
\maketitle

\begin{abstract}
We describe simple linear time algorithms for coloring the squares of
balanced and unbalanced quadtrees so that no two adjacent squares
are given the same color.  If squares sharing sides are defined as
adjacent, we color balanced quadtrees with three colors, and unbalanced
quadtrees with four colors; these results are both tight, as some
quadtrees require this many colors.  If squares sharing corners are
defined as adjacent, we color balanced or unbalanced quadtrees with six
colors; for some quadtrees, at least five colors are required.
\end{abstract}

\section{Introduction}

A {\em quadtree} \cite{Samet} is a data structure formed by starting from
a single square, and recursively dividing squares into four smaller
squares.  In this paper we consider problems of coloring quadtree
squares so that no two neighboring squares have the same color.  This
quadtree coloring problem was introduced by Benantar et al
\cite{BDFK,BFK}, motivated by problems of scheduling parallel
computations on quadtree-structured finite element meshes.

There are several variants of the problem depending on the details of
its definition. Quadtrees may be {\em balanced} (i.e.
squares sharing an edge may be required to be within a factor of two of
each other in size) or {\em unbalanced}.  Balanced quadtrees are
typically used in finite element meshes, but other applications may give
rise to unbalanced quadtrees. Further, squares may be defined to be
neighboring if they share a portion of an edge ({\em edge adjacency}), or
if they share any vertex or portion of an edge ({\em vertex adjacency}).
We can thus distinguish four variants of the problem: balanced with edge
adjacency, unbalanced with edge adjacency, balanced with corner
adjacency, and unbalanced with corner adjacency. (Other balance
conditions may also be used, but we do not concern ourselves with them
here.)

Since quadtrees are planar,
the four-color theorem for planar maps implies that edge-adjacent
quadtrees require at most four colors, regardless of balance. Benantar et
al. showed that with corner adjacency, balanced quadtrees require at
most six colors \cite{BFK} and unbalanced quadtrees require at most
eight colors \cite{BDFK}. Benantar et al also suggest that four colors
may suffice, even for corner adjacency \cite{BDFK}.

Here, we tighten the upper bounds above, and show that
balanced edge-adjacent quadtrees require only three colors while even
unbalanced corner-adjacent quadtrees can be six-colored.
We provide simple linear time algorithms that color quadtrees within
these bounds, and that four-color edge-adjacent unbalanced quadtrees.
We also provide lower bound examples showing that three
colors are necessary for balanced edge adjacency, four colors are
necessary for unbalanced edge adjacency, and at least five colors are
necessary for balanced corner adjacency, refuting the suggested
four-color bound of Benantar et~al.

\section{Balanced edge adjacency}

\begin{figure}[t]
$$\efig{2in}{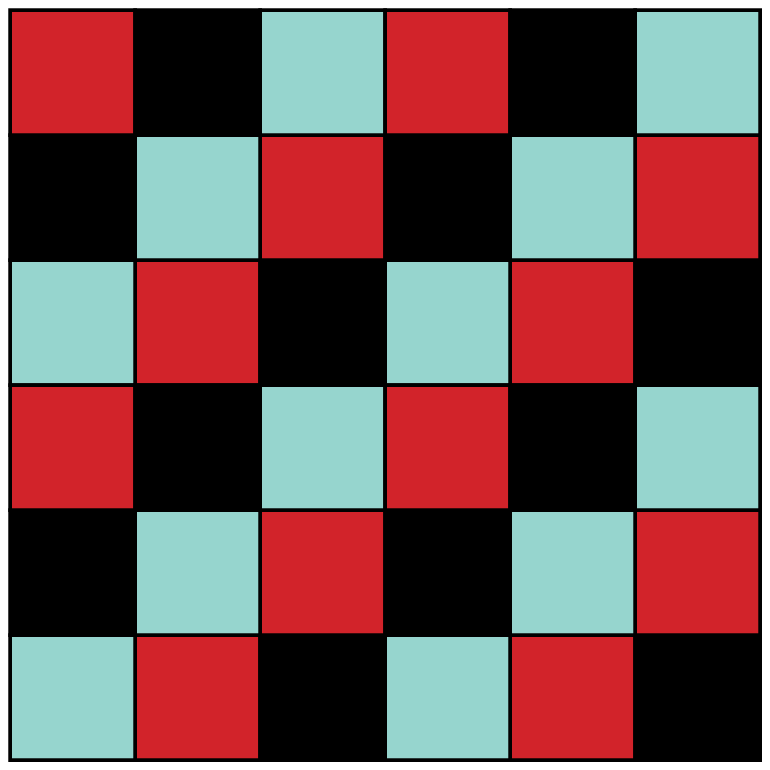}\qquad\efig{2in}{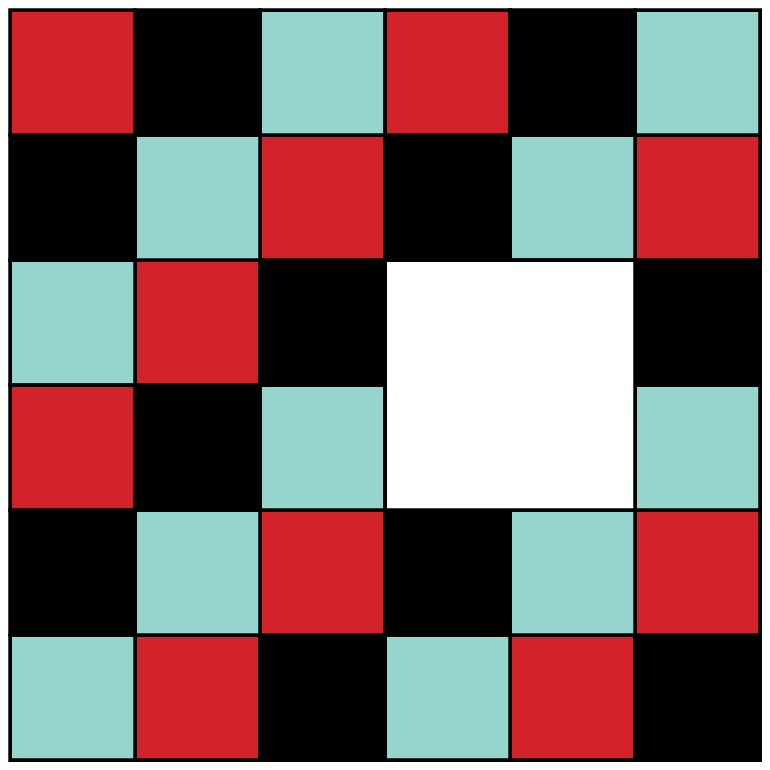}$$
\caption{(a) Three-coloring of grid; (b) Any four-square hole has only
two neighboring colors.}
\label{grid}
\end{figure}

\begin{theorem}\label{three}
Any balanced quadtree
can be colored with three colors so that no two squares
sharing an edge have the same color.
\end{theorem}

\begin{proof}
Imagine constructing the quadtree bottom-up, by starting with a regular
grid of squares and then consolidating quadruples of squares of one size
to make squares of the next larger size.  We color the initial grid by a
regular pattern of three colors, depicted in Figure~\ref{grid}(a).
Then, when we consolidate four squares of one size to make squares of
the next larger size, each larger square has only two colors among its
smaller neighbors (Figure~\ref{grid}(b)), forcing it to take the third
color.  Connected sets of larger squares then end up colored by the same
regular pattern used to color the smaller grid, so we can repeat this
process of consolidation and coloring within each such set.
\end{proof}

We note that this process gives each square a color depending only on
its size and position within the quadtree, and not depending on what
subdivisions have occurred elsewhere in the quadtree.  This coloring can
be determined easily from the color the square's parent would be given
by the same process, so the coloring algorithm can be performed top-down
in linear time.

\section{Unbalanced edge adjacency}

\begin{figure}[t]
$$\efig{3.5in}{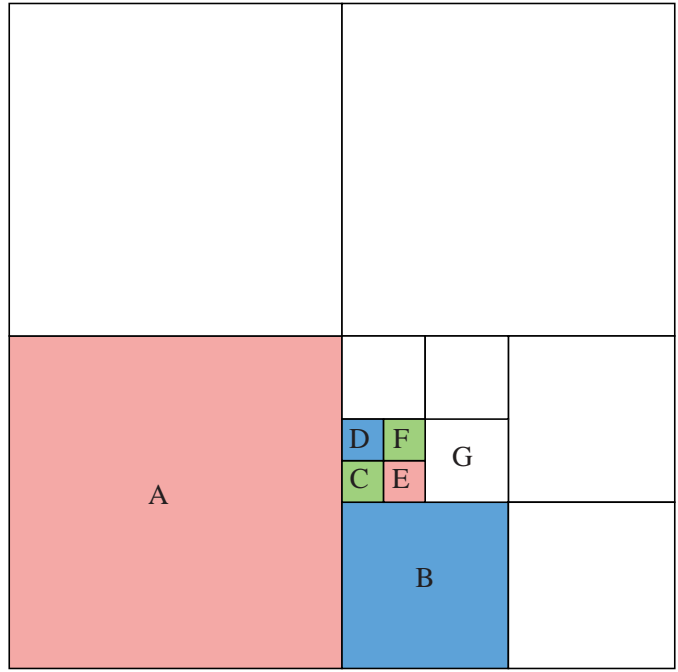}$$
\caption{Unbalanced edge adjacency requires four colors.}
\label{unbal-4x}
\end{figure}

By the four-color theorem for planar maps, any unbalanced
quadtree can be colored with four colors so that no two
squares sharing an edge have the same color.  Such a coloring is not
difficult to find:

\begin{theorem}\label{four}
Any unbalanced quadtree
can be colored in linear time with four colors so that no two squares
sharing an edge have the same color.
\end{theorem}

\begin{proof}
We form the desired quadtree by splitting squares one at a time; at each
step we split the largest square possible.  Thus the four smaller
squares formed by each split are, at the time of the split, among the
smallest squares in the quadtree.  As we perform this splitting process,
we maintain a valid four-coloring of the quadtree.

When we split a square, we color the four resulting smaller squares.  We
give the upper right and lower left squares the same color as their
parent.  Each of the other two squares has at most four neighbors,
two of which are the same color.  Therefore each has at most three
neighboring colors, and at least one color remains available; we give each
of these two squares one of the available colors. 
\end{proof}

As we now show, four colors may sometimes be necessary.

\begin{theorem}
There is an unbalanced quadtree requiring four colors for all colorings in
which no two squares sharing an edge have the same color.
\end{theorem}

\begin{proof}
An unbalanced quadtree is depicted in Figure~\ref{unbal-4x}, with some
of its squares labeled. A simple case argument shows that it has no
three-coloring: suppose for a contradiction that we are attempting to
color it red, blue, and green.  Since squares $A$, $B$, and $C$ are
mutually adjacent, we may assume without loss of generality that they
are colored red, blue, and green respectively.  Since $D$ is adjacent to
$A$ and $C$, it must be blue, and since $E$ is adjacent to $B$ and $C$,
it must be red.  Since $F$ is adjacent to $D$ and $E$, it must be green.
But then $G$ is adjacent to a red square ($E$), a green square ($F$),
and a blue square ($B$), so it can not be given any of the three colors.
Thus, four colors are required to color this quadtree.
\end{proof}

\section{Balanced corner adjacency}

\begin{figure}[t]
$$\efig{3.5in}{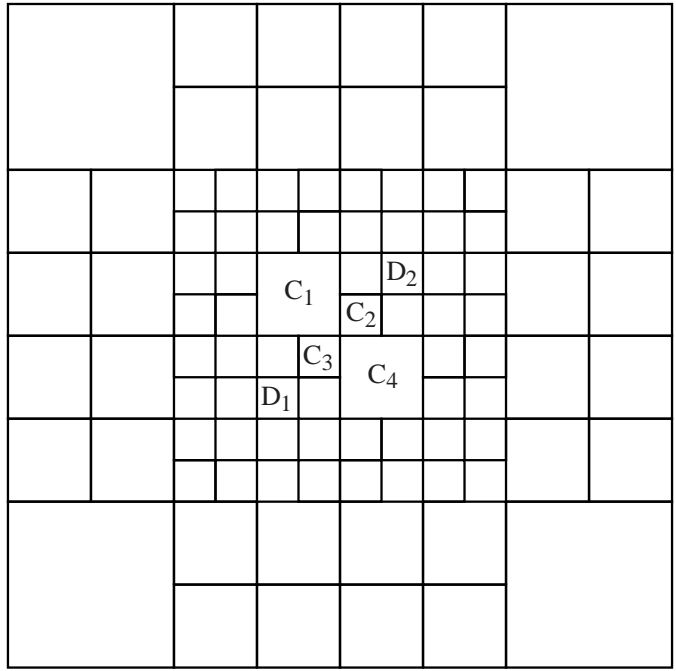}$$
\caption{Balanced corner adjacency requires at least five colors.}
\label{bal-5x}
\end{figure}

\begin{figure}[p]
$$\efig{3.5in}{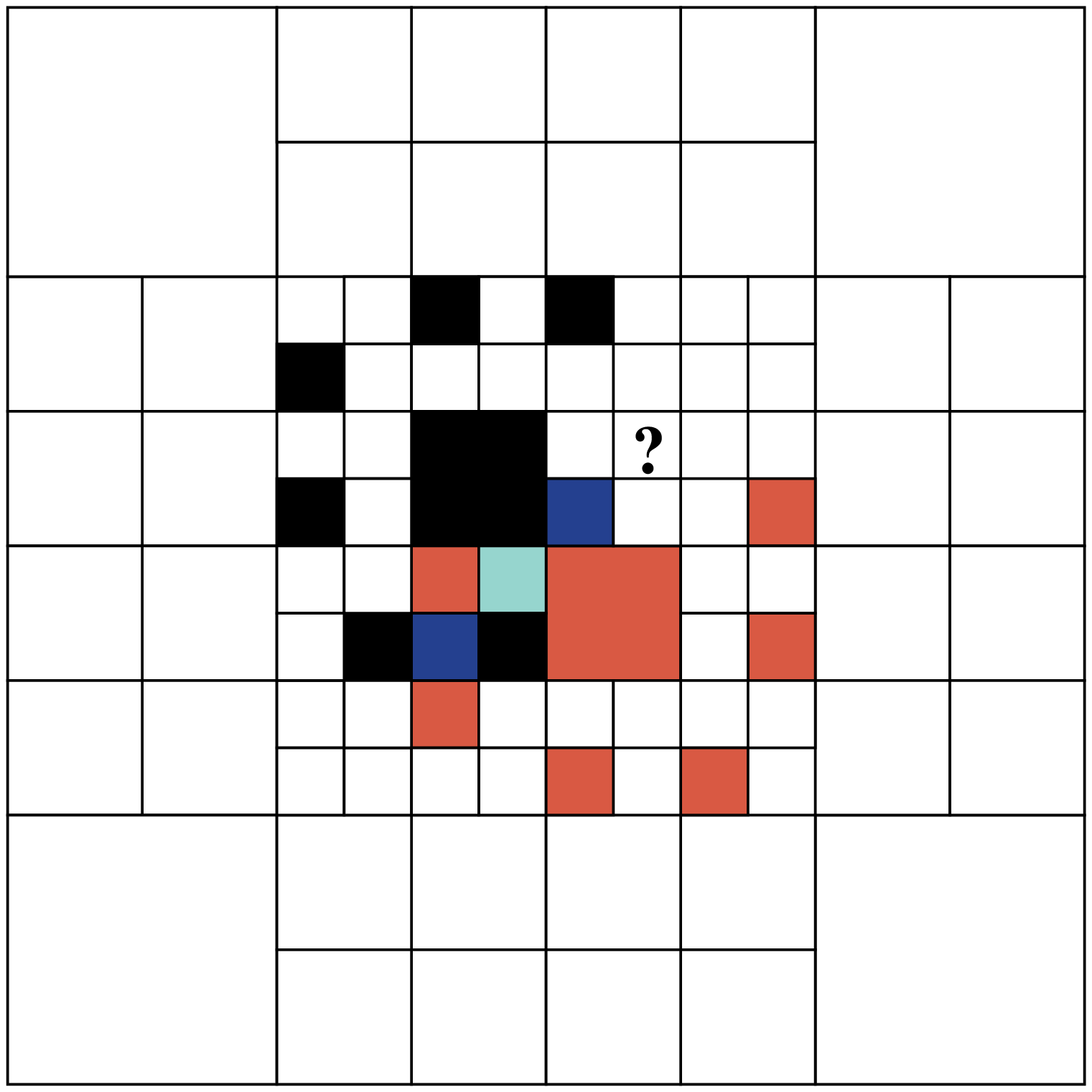}$$
\caption{Forced squares after choosing colors of center squares, with
neighboring square $D_1$ colored the same as a small center square.}
\label{4color1}
\end{figure}

\begin{figure}[p]
$$\efig{3.5in}{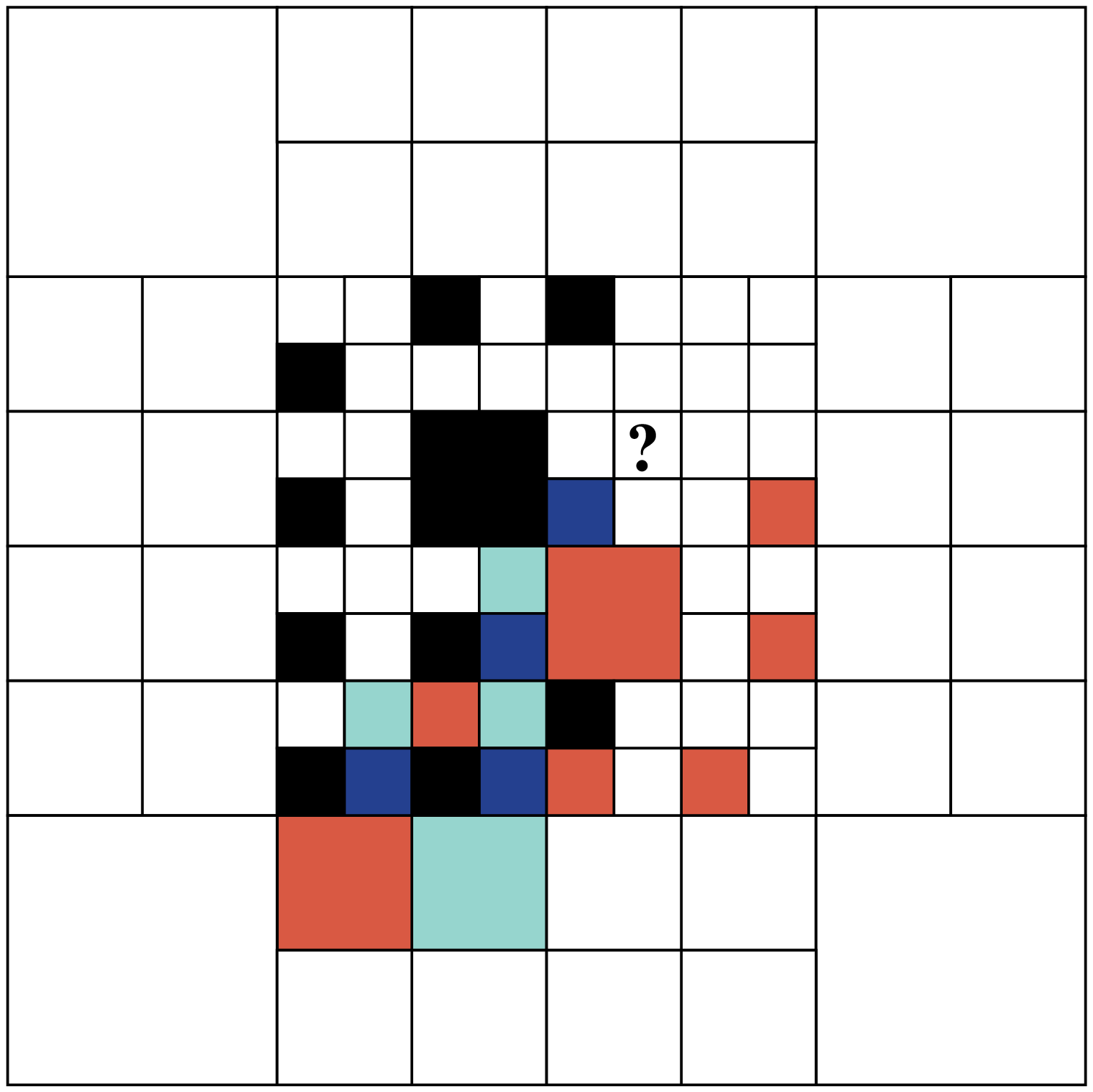}$$
\caption{Forced squares after choosing colors of center squares, with
neighboring square $D_1$ colored the same as a large center square.}
\label{4color2}
\end{figure}

\begin{theorem}
There is a balanced quadtree requiring five colors for all colorings in
which no two squares sharing an edge or a corner have the same color.
\end{theorem}

\begin{proof}
A balanced quadtree is depicted in Figure~\ref{bal-5x}.
A simple case argument shows that it has no four-coloring:
choose four different colors for the four squares $C_1$, $C_2$,
$C_3$, and $C_4$ meeting in the center vertex.  Then, choose a color for
one of the diagonal neighbors, $D_1$ and $D_2$, of the two small center
squares.  Now repeatedly apply the following two coloring rules:
\begin{enumerate}
\item If some square $s$ has three differently colored neighbors, assign
the remaining fourth color to~$s$.
\item If some square
$s$ has a corner shared by three other squares, each of which is adjacent
to squares of some color
$a$, assign color $a$ to $s$ since no other choice leaves enough free
colors to the other squares sharing the corner.
\end{enumerate}
Figures \ref{4color1} and~\ref{4color2} show the results of a partial
application of these rules, for two choices of color for $D_1$.  The
third possible choice is symmetric with Figure~\ref{4color2}. No matter
what color is chosen for
$D_1$, these rules lead to an inconsistency at $D_2$: rule
2 applies in two different ways, forcing $D_2$ to have two different
colors.  Therefore the overall quadtree can not be colored.
\end{proof}

\section{Unbalanced corner adjacency}

\begin{theorem}\label{six}
Any balanced or unbalanced quadtree
can be colored in linear time with six colors so that no two squares
sharing an edge or a corner have the same color.
\end{theorem}

\begin{proof}
We form the adjacency graph of the squares in the quadtree, and apply
the {\em greedy algorithm}: remove a minimum degree vertex from the
graph, color recursively, then add back the removed vertex and give it a
color different from its neighbors.  If the maximum degree of a vertex
removed at any step is $d$, this uses at most $d+1$ colors.
We can find the minimum degree vertex by
maintaining for each $i\le 5$ a doubly linked lists of the vertices
currently having degree $i$; as we show below, at least one list will be
nonempty, and it is straightforward to update these lists in constant
time per step.  Therefore, the overall time will be linear.

Our bound
of six colors then follows from the following lemma.
Let $Q$ be a subset of the squares in a (not-necessarily balanced)
quadtree.  Define a {\em big box} to be a square that is not the smallest
in $Q$, that has at most five neighbors which are also not the smallest in
$Q$ (Figure~\ref{configs}(a)). Define a {\em hanging box} to be a square
$s$ that is not the smallest in $Q$, that has at most three neighbors
incident to the upper left corner, and at most two below or to the right;
the below-right neighbors must also not be the smallest in $Q$
(Figure~\ref{configs}(b)).

Define a {\em good chain} to be a set of one or more squares
all the smallest in $Q$, with the following properties
(Figure~\ref{configs}(c)): Each square in the chain must have at most one
neighbor below it; except for the bottommost square in the chain, this
neighbor must be another square in the chain, adjacent at the bottom left
corner. The bottommost square in the chain can be adjacent to a square
$s$ below it and outside the chain, but only if $s$ is larger than the
squares in the chain.  Similarly, each square in the chain must have at
most one neighbor to the right of it; except for the topmost square in
the chain, this neighbor must be another square in the chain, adjacent at
the top right corner.  The topmost square in the chain can be adjacent to
a square $s$ to the right of it and outside the chain, but again only if
$s$ is larger than the squares in the chain.
If the chain has exactly one square in it, it may have neighbors both
below and to the right, as long as both neighbors are larger.

Finally, define a {\em good configuration} to be any one of these three
patterns: a big box, a hanging box, or a good chain.
Note that all three of these configurations give a degree-five
square or squares.

\begin{figure}[t]
$$\efig{1.75in}{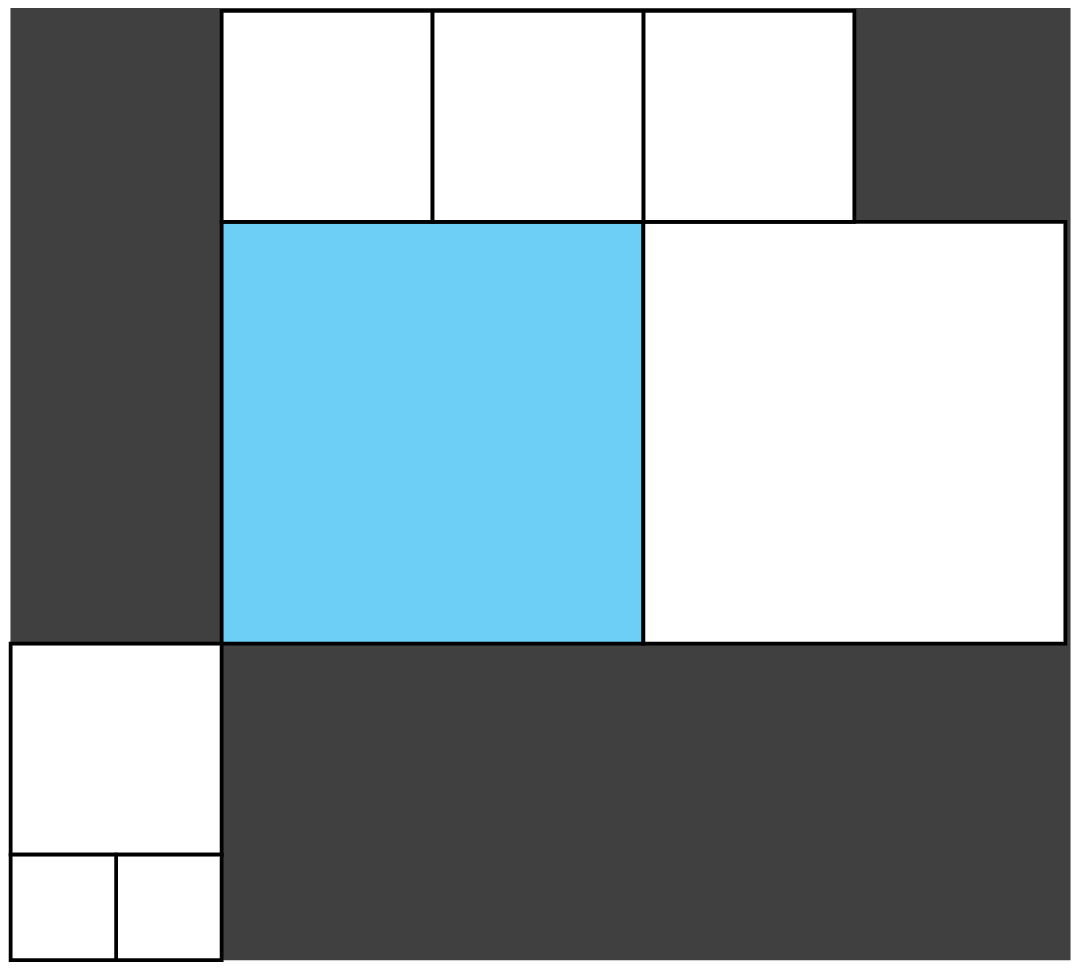}
\qquad\efig{1.75in}{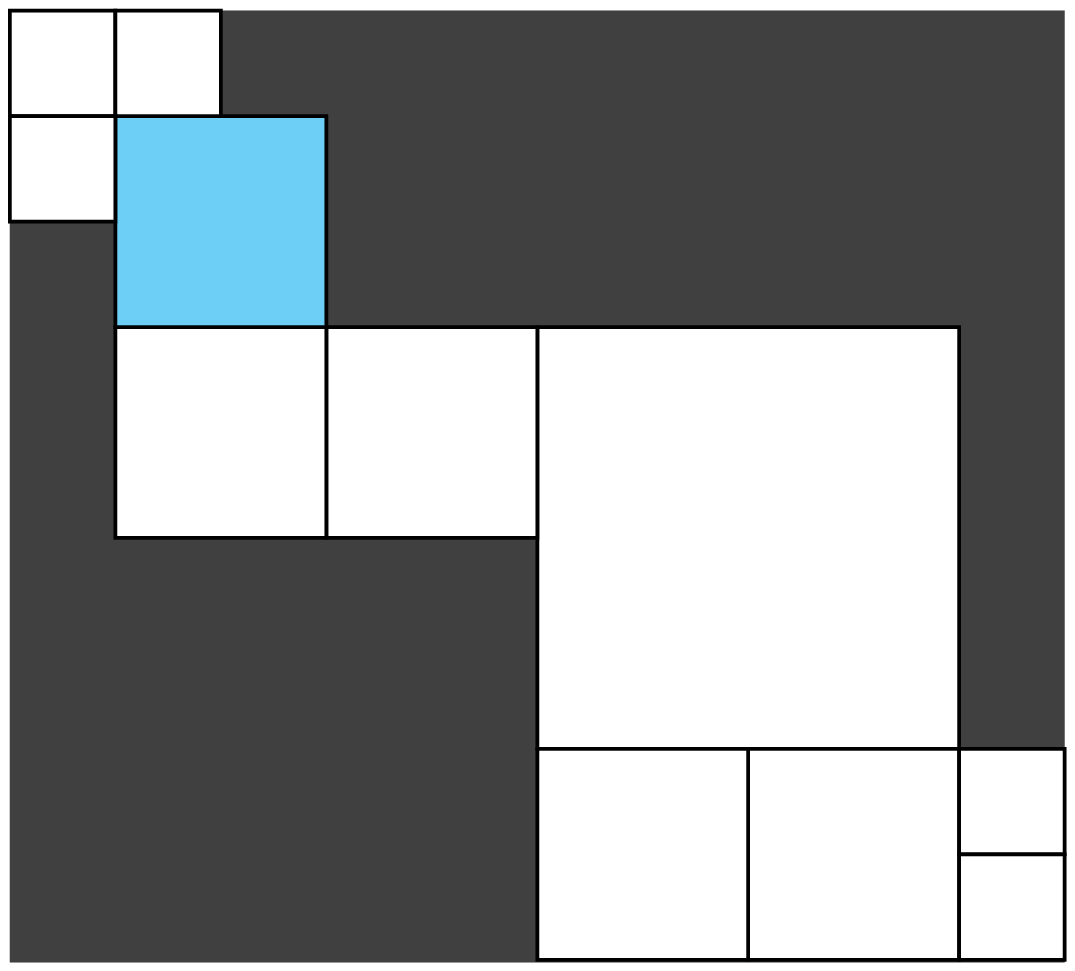}$$
\vspace{0.1in}
$$\efig{1.75in}{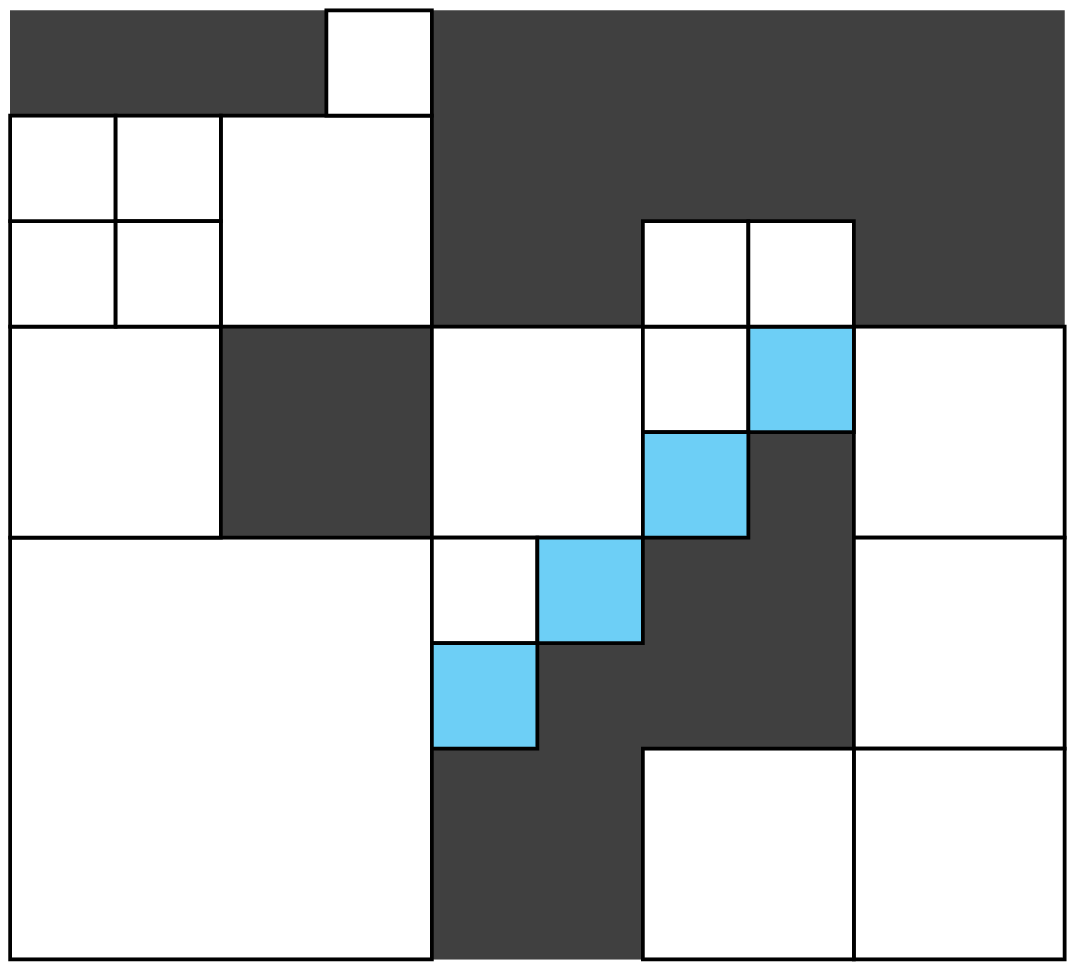}$$
\caption{Good configurations: In each figure, the light and shaded
squares represent a subset of a quadtree, and the shaded squares form a
good configuration for that subset. (a) big box; (b) hanging box; (c) good
chain.}
\label{configs}
\end{figure}

\begin{lemma}
Let $Q$ be any subset of the squares of a quadtree.  Then $Q$ has a good
configuration.
\end{lemma}

\begin{proof}
We use induction on the number of levels in $Q$.
Let $Q'$ be formed by replacing each smallest square in Q by its parent.
(We think of $Q$ as being formed by splitting some squares in $Q'$
and removing some of the resulting children.)
Let $C$ be a good configuration in $Q'$.

First, suppose $C$ is a big box in $Q'$.  Then it is also a big box in
$Q$ since none of its neighbors can be subdivided.

Next, suppose $C$ is a hanging box in $Q'$.  If none of its neighbors is
subdivided to form $Q$, it is a big box in $Q$.  If one of its neighbors
is subdivided and has a child neighboring $C$ and not incident to the
upper left corner of $C$, that child is a (singleton) good chain (its only
below-right adjacency is to $C$ itself).  If $C$'s neighbors are
subdivided but the only children neighboring $C$ are on the corner, $C$
remains a hanging box in $Q$.

Finally, suppose $C$ is a good chain in $Q'$.  If some square of $C$ is
subdivided, and its lower right child is in $Q$, that child is a
(singleton) good chain in $Q$.  If not, but some squares are subdivided
and have upper right or lower left children, any maximal contiguous
sequence of such children is a good chain in $Q$.  If neither of these
two cases holds, but some squares are subdivided and have only their upper
left children in
$Q$, then some sequence of such children and of lower right children of
neighbors of $C$ forms a good chain in $Q$.  If no squares in $C$ are
subdivided and none of their upper or left neighbors are subdivided, each
square in the chain becomes a big box in $Q$.  If no squares in $C$
are subdivided, some upper or left neighbor is subdivided, and its lower
right child is in $Q$, that child is a singleton good chain.  In the
remaining case, any subdivided neighbor has neighboring children only on
the upper left corners of squares in $C$, and all squares in $C$ become
hanging boxes in
$Q$.
\end{proof}

By the lemma above, any graph formed by a subset of the
quadtree squares has a vertex of degree at most five, so the greedy
algorithm uses at most six colors. This concludes the proof of
Theorem~\ref{six}.
\end{proof}

\section{Conclusions}

We have shown that balanced edge-adjacent quadtrees require three
colors, and unbalanced edge-adjacent quadtrees require four colors.
Corner-adjacent quadtrees may require either five or six colors.
It remains to close this gap in the corner-adjacent case and to
determine whether the balance condition makes a difference in this case.

\end{document}